\documentclass{emulateapj}
\usepackage{epsfig}

\newcommand{\etal}{\mbox{et al.}}
\newcommand{\ergcms}{erg cm$^{-2}$ s$^{-1}$}
\newcommand{\ergs}{erg s$^{-1}$}
\newcommand{\phcms}{ph cm$^{-2}$ s$^{-1}$}
\newcommand{\degree}{$^\circ$}

\newcommand{\chandra}{{\it Chandra}}

\newcommand{\sgrastar}{\mbox{Sgr A$^*$}}

\shortauthors{Muno \etal}
\shorttitle{Iron Reflection Nebulae}

\begin{document}

\title{Discovery of Variable Iron Fluorescence from Reflection Nebulae in the Galactic Center}

\author{
M. P. Muno\altaffilmark{1}, F. K. Baganoff\altaffilmark{2}, W. N. Brandt,\altaffilmark{3}, S. Park,\altaffilmark{3} \& M. R. Morris\altaffilmark{4}}

\altaffiltext{1}{Space Radiation Laboratory, California Institute of Technology, Pasadena, CA 91125; mmuno@srl.caltech.edu}
\altaffiltext{2}{Kavli Institute for Astrophysics and Space Research,
Massachusetts Institute of Technology, Cambridge, MA 02139; fkb@space.mit.edu}
\altaffiltext{3}{Department of Astronomy and Astrophysics, 
The Pennsylvania State University, University Park, PA 16802; niel,park@astro.psu.edu}
\altaffiltext{4}{Department of Physics and Astronomy, University of California,
Los Angeles, CA 90095; morris@astro.ucla.edu}

\begin{abstract}
Based on three years of deep observations of the Galactic center with
the \chandra\ X-ray Observatory, we report the discovery of changes in
the intensities and morphologies of two hard X-ray nebulosities. The
nebulosities are dominated by fluorescent iron emission, and are
coincident with molecular clouds. The morphological changes are manifest 
on parsec scales, which requires that these iron features are scattered 
X-rays from a 2 or 3-year-long outburst of a point source 
(either \sgrastar\ or an X-ray binary) with a luminosity of at least 
$10^{37}$ \ergs.  The variability precludes the
hypotheses that these nebulae either are produced by keV electrons
bombarding molecular clouds, or are iron-rich ejecta from
supernovae.  Moreover, the morphologies of the reflection nebulae implies
that the dense regions of the clouds are filamentary, with
widths of $\approx$0.3 pc and lengths of $\approx$ 2 pc.
\end{abstract}


\section{Introduction}

As a result of the high density of matter and stars in the Galactic
center, opportunities arise for energetic particles, radiation, and 
interstellar matter to 
interact in ways that occur very rarely in the Galactic disk. One 
manifestation of these interactions are iron fluorescence nebulae 
that are visible in the X-ray band, and that are 
associated with 
the molecular cloud complexes Sgr B1 and B2 \citep{mur00,mur01b}, 
Sgr C \citep{mur01a}, and M 0.11-0.11 \citep{par04}. The fluorescence 
is thought to be produced when 
low-ionization iron in the molecular clouds is bombarded either by 
$\sim$30 keV electrons \citep{val00,ylw02,byk02} or by $\sim$8 keV X-rays 
\citep{sc98,rev04}. However, exciting the fluorescent features 
with electrons would require a large, and arguably unreasonable, amount of 
energy in keV electrons \citep{rev04}, and there is currently no X-ray 
source in the Galactic center bright enough to illuminate 
these features \citep{smp93,koy96,rev04b}. Therefore, the excitation mechanism
for these features is uncertain.

In this paper, we attempt to determine how the iron fluorescence features 
are excited, by searching for variations in the morphologies and intensities
of the two examples with the highest surface brightnesses 
\citep[Features 1 and 2 in][]{par04}, which are located in the M0.11-0.11 cloud
complex. If the fluorescence is excited by particles, which would 
propagate rather slowly ($<$$0.06 c$ for 30 keV electrons), the 
features would not vary over the course of a few years. Alternatively, if the
fluorescence is produced by X-rays, the absence of any source that is
currently bright enough to produce them implies that the
X-rays were produced in a transient event. If the outburst
were short enough, then in a few years the light front could propagate
across the features, causing their morphologies and intensities to
change \citep{sc98}. Here we report that, in fact, the two fluorescent 
features have changed after three years.

\begin{figure} 
\centerline{\psfig{file=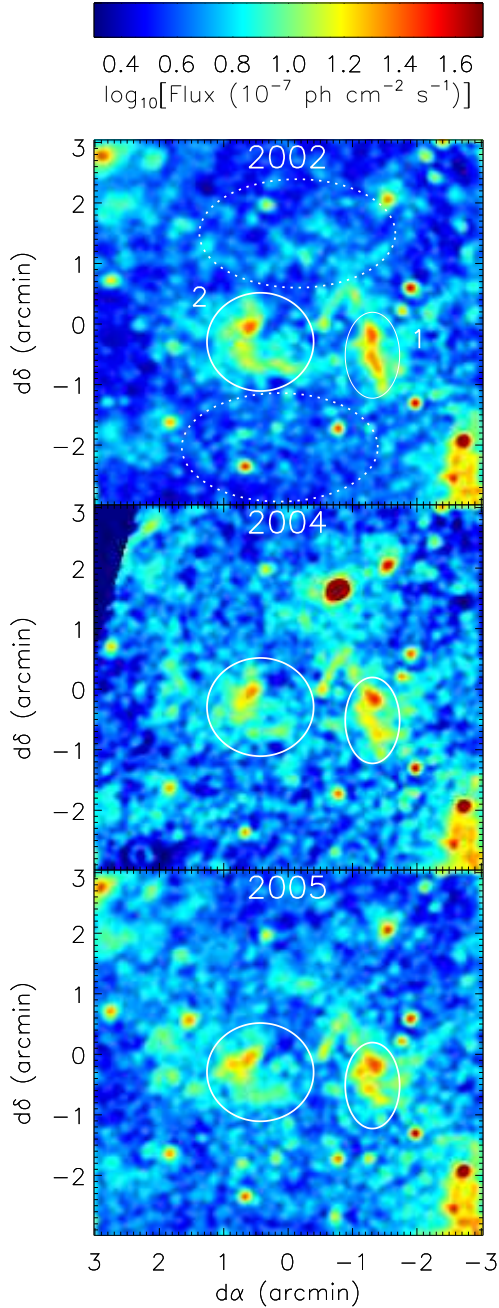,width=0.8\linewidth}}
\caption{
Hard (4--8 keV) X-ray images of the 6\arcmin $\times$ 6\arcmin\ field
around the iron fluorescent features (centered at 
$\alpha$, $\delta$ = 266.44156, --28.94267 [J2000]), illustrating 
their evolution as 
a function of time. The solid ellipses denote the locations of the fluorescent
iron features from \citet{par04}. The dashed ellipses in the top
panel denote the regions used for background extraction. 
The bar at top displays the mapping between flux and color, with red
indicating the brightest emission, and blue the faintest.
}
\label{fig:images}
\end{figure}

\begin{figure}
\centerline{\psfig{file=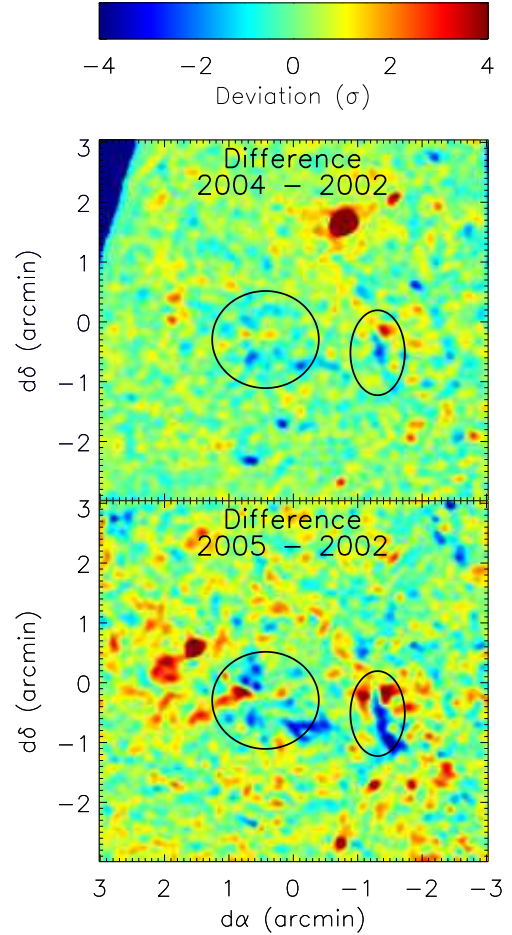,width=0.8\linewidth}}
\caption{
Significance maps of the difference in flux received in the 4--8 keV 
band. Red indicates an increase in intensity. At the locations of the 
fluorescent features, the variations in flux are spatially coherent, indicating
that the diffuse features have changed in shape and intensity.
The changes are more significant in the 2005--2002 difference map, because
the exposure was twice that in 2004. The bar at the top of the figure provides
the mapping between significance and color.}
\label{fig:difference}
\end{figure}

\section{Observations}

The features that we studied were identified by
\citet{par04}, based on 500 ks of observations of the 20 pc
around \sgrastar\ taken with the Advanced CCD Imaging Spectrometer
imaging array 
between 2002 May 22 and June
3. Since then, 100 ks of data were acquired between 2004 July 5 and 6,
and 200 ks between 2005 July 24 and August 1. All of the
observations had aim points within 20\arcsec\ of \sgrastar, although
the roll angles for the satellite differed in
2002 (76\degree) and 2004-2005 (280\degree). Features 1 and 2 are
located near $\alpha$=266.492\degree,
$\delta$=--28.943\degree\ (J2000), or 6\arcmin\ from
\sgrastar, and are covered by all of our observations.\footnote{A
third feature was also reported at $\alpha$=266.593, $\delta$=--28.920
in \citet{par04}, but was not covered by the observations in
2004--2005, and so we ignore this feature.} At this offset, the point-spread 
function (PSF) at 6 keV has a full-width at half-maximum (FWHM) of 6\arcsec, 
which corresponds to a physical scale of 0.2 pc \citep[$D$=8 kpc;][]{mcn00}.

We processed the event lists from each observation to remove bad pixels, 
cosmic rays, and background flares in the standard
manner\footnote{http://asc.harvard.edu/ciao/threads/}, using CIAO
version 3.3.0 and the calibration database version 3.2.2.  Next, we 
identified point-like X-ray sources using the wavelet algorithm
wavdetect. Finally, the absolute astrometric frame was registered
using X-ray sources with counterparts in the 2MASS catalog \citep{skr06},
making it accurate to 0\farcs2.

To search for temporal evolution, we 
produced three separate images of the flux in the 4--8 keV band, using
the data from 2002, 2004, and 2005. This limited 
energy range was chosen to reject foreground emission, emission from 
soft plasma, and particle background. In Figure~\ref{fig:images}, we 
display the portion of the field containing Features 1 and 2, 
smoothed by a Gaussian with a FWHM of 6\arcsec. Having smoothed the images 
on this scale, changes in the PSF shape are not significant, as can be seen by 
examining point sources in the image. Over 
three years, the morphologies of Features 1 and 2 change noticeably. 

These variations can be seen more clearly in difference images, which 
we have normalized to the uncertainty in the convolved flux maps and displayed
in Figure \ref{fig:difference}. To generate these, we first
calculated the one-sided, upper 1$\sigma$ Poisson uncertainty on 
the counts in each pixel of each image, following \citet{geh86}. We 
divided this by the corresponding exposure maps to obtain the uncertainties
on the flux, $\sigma_{f,i}$. The uncertainties in each convolved image 
($\sigma_{C,i}$) were 
then the square root of the convolution of the square of the uncertainties
with the square of the 
Gaussian smoothing function ($g$), 
$\sigma_{C,i} = (\sigma_{f}^2 * g^2)_i^{1/2}$.
The uncertainties in each pair of images were then added in quadrature to
obtain those of the difference images. The difference images
were divided by the uncertainty maps to obtain the significance maps. 

In Figure~\ref{fig:difference}, red features were defined to be brighter 
in the later images; a clear example is the transient source 
XMM J174554.4-285456 \citep{por05,m-trans}, which appeared in 2004
just north of the center of the image.  We find significant variations 
in several places in the field. Some of the most prominent variations are
at the locations of the two fluorescent features with the highest surface
brightness. The main ridge that
composed Feature 1 in 2002 faded at the 3$\sigma$ level by 2005, while the
northern part of the feature on either side of the ridge
brightened by 4$\sigma$. For Feature 2, we find that the southern portion 
faded by 3$\sigma$ between 2002 and 2005, whereas the eastern portion 
shifted farther east and became more extended at the 4--5$\sigma$ level.
We also find 3--4$\sigma$ increases in the flux to the east of Feature 2 
and the south of Feature 1. The pervasive nature of these variations
is not surprising, becuase then entire field contains molecular gas 
\citep{thu99} and exhibits diffuse iron K$\alpha$ emission \citep{m-diff}.

\begin{figure}
\centerline{\psfig{file=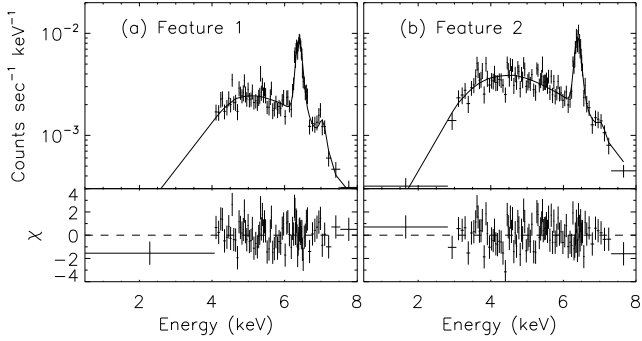,width=\linewidth}}
\caption{
Spectra of fluorescent Features 1 and 2. The top panels display the
spectra in units of detector counts (crosses). The solid line is the 
best-fit model of an absorbed power law and two narrow lines at 
6.4 and 7.0 keV. The bottom panels display the residuals, in units of 
the statistical uncertainty in the counts in each bin. The dominant 
feature in each spectrum is the iron line at 6.4 keV.
}
\label{fig:spectra}
\end{figure}

Next, we extracted spectra of the diffuse features. The source
spectra were extracted from the solid ellipses in Figure~\ref{fig:images}
(top panel)
and the background from the dotted ellipses above and below 
the features. We 
excluded events that fell within circles circumscribing the 92\% 
encircled-energy contours of the PSFs for each point source.  
The background-subtracted spectra were grouped so that each bin had a 
signal-to-noise ratio of $>$5. Spectra for all of the data are displayed in 
Figure~\ref{fig:spectra}. 

We modeled the spectra from the 2002 and 2004--2005 epochs jointly
using XSPEC, 
as a power-law continuum with photon index $\Gamma$ and
normalization $N_{\Gamma}$, two narrow iron lines at 
6.40 (Fe K$\alpha$) and 7.06 keV (Fe K$\beta$). We assumed that 
this emission was absorbed by the interstellar medium ($N_{\rm H, ISM})$, 
plus a component of the absorption intrinsic to the associated molecular 
clouds ($N_{\rm H,cloud}$) that covered a variable fraction of the 
emitting region ($f_{\rm cloud}$). The use of such a ``partial covering'' 
factor ($f_{\rm cloud}$) was motivated by the assumption 
that the emission from the clouds would take place at a range of optical 
depths. We also assumed that $\Gamma$ and $N_{\rm H, ISM}$ 
were identical for both features during both epochs.
In Table~\ref{tab:spec}, we list the best-fit parameter values, 
iron equivalent widths, and fluxes.

\begin{deluxetable}{lcccc}
\tablecolumns{5}
\tablewidth{0pc}
\tablecaption{Spectra of the Iron Fluorescence Features\label{tab:spec}}
\tablehead{
\colhead{} & \multicolumn{2}{c}{Feature 1} & \multicolumn{2}{c}{Feature 2}\\
\colhead{} & \colhead{2002} & \colhead{2004--2005} & \colhead{2002} & \colhead{2004--2005}
} 
\startdata
$N_{\rm H,ISM}$ & $11_{-1}^{+2}$ & 11\tablenotemark{a} & 11\tablenotemark{a} & 11\tablenotemark{a} \\ 
$N_{\rm H,cloud}$ & $25_{-2}^{+4}$ & $23_{-2}^{+4}$ & $29_{-6}^{+20}$ & $64_{-18}^{+26}$ \\
$f_{\rm cloud}$ & 1.0\tablenotemark{b} & 1.0\tablenotemark{b} & $0.61_{-0.05}^{+0.10}$ & $0.76_{-0.05}^{+0.08}$ \\
$\Gamma$ &  $1.84_{-0.13}^{+0.03}$ & 1.84\tablenotemark{a} & 1.84\tablenotemark{a} & 1.84\tablenotemark{a} \\
$N_\Gamma$ & $6_{-2}^{+2}$ & $5_{-2}^{+4}$ & $6_{-2}^{+1}$ & $8_{-2}^{+6}$ \\
$F_{K\alpha}$ & $18_{-1}^{+2}$ & $16_{-1}^{+1}$ & $19_{-2}^{+3}$ & $18_{-4}^{+8}$ \\
$F_{K\beta}$ &  $2.1_{-0.6}^{+0.6}$ & $1.9_{-0.9}^{+0.9}$ & $1.5_{-0.8}^{+0.8}$ & $1.9_{-1.8}^{+1.7}$ \\ [3pt]
EW$_{K\alpha}$ & $1000_{-90}^{+240}$ & $1010_{-90}^{+190}$ & $930_{-160}^{+160}$ & $690_{-220}^{+220}$ \\ 
EW$_{K\beta}$ & $140_{-20}^{+70}$ & $150_{-70}^{+60}$ & $90_{-45}^{+40}$ & $90_{-80}^{+100}$ \\
$F_{\rm X}$ & $4.0_{-0.2}^{+0.2}$ & $3.7_{-0.2}^{+0.2}$ & $5.7_{-0.3}^{+0.3}$ & $4.2_{-0.3}^{+0.3}$ \\
$F^\prime_{\rm X}$ & $5.2_{-0.3}^{+0.2}$ & $4.8_{-0.3}^{+0.2}$ & $7.7_{-0.4}^{+0.5}$ & $5.5_{-0.5}^{+0.3}$
\enddata
\tablenotetext{a}{The interstellar absorption and photon index of the 
continuum were assumed to be identical for each spectrum.}
\tablenotetext{b}{For Feature 1, $f_{\rm cloud}$ was poorly-constrained, 
so we fixed the parameter to its best-fit value of 1.}
\tablecomments{The reduced chi-squared for the joint fit was 211/176.
The units are as follows: $N_{\rm H,ISM}$ and $N_{\rm H,cloud}$ are 
$10^{22}$ cm$^{-2}$; $N_\Gamma$ is $10^{-4}$ \phcms\ keV$^{-1}$ at 1 keV; 
$F_{K\alpha}$ and $F_{K\beta}$ are $10^{-6}$ \phcms;
EW$_{K\alpha}$ and EW$_{K\beta}$ are eV; $F_{\rm X}$ is the observed 4--8 keV 
flux in units of $10^{-13}$ \ergcms; and $F^\prime_{\rm X}$ 
is the absorption-corrected 4--8 keV flux in units of $10^{-13}$ \ergcms.
Uncertainties are 1$\sigma$, for 1 degree of freedom for the parameters, and 
for 7 degrees of freedom for the fluxes and equivalent widths.}
\end{deluxetable}

As found by \citet{par04}, both features have extremely strong Fe
K$\alpha$ emission, with equivalent widths of $\approx$1 keV. The Fe
K$\beta$ emission is $\approx$10-15\% of the strength of the
K$\alpha$ line, as is expected for fluorescent emission
\citep[e.g.,][]{mur00}. Both features are heavily absorbed, so
most of the observed flux is received between 4 and 8 keV.
We find that for Feature 1, despite the noticeable changes in the
morphology, the 4--8 keV flux and the iron flux in the extraction
region remained constant between 2002 and 2005. For
Feature 2, the inferred iron flux remained constant, but the observed 
4--8 keV flux declined by 26$\pm$7\%. 

\section{Discussion}

We have found that two fluorescent iron features located 6\arcmin\
(14 pc in projection) from \sgrastar\ 
varied in morphology and intensity between 2002 and
2005 (Figs.~\ref{fig:images} and \ref{fig:difference}, and
Tab.~\ref{tab:spec}). These changes in the
fluorescence features require that the illuminating source varied on
a time scale shorter than 3 years. Moreover, the features are
$\approx$0.3\arcmin\ wide and $\approx$ 1\arcmin\ long, which
corresponds to projected physical sizes of $\approx$0.7 $\times$ 2 pc. The 
smaller of these dimensions constrains the minimum duration of
the outburst to have been two years. The fact that two features 
with a projected spearation of $\approx$12 light-years varied over 
the same time interval implies that the illuminating 
source and the target clouds have separations with a component along
our line of sight.

Such rapid changes over the course of a few years would not be produced by
steady, slow-moving sources such as ejecta from supernovae
\citep{byk02}.  These shifting iron fluorescence features also would not 
be produced by $\approx$30~keV electrons, which would have velocities of
only 0.06$c$ (and slower diffusion speeds), and which would presumably
be accelerated in shocks that persist for much longer than a few years.
The most reasonable explanation for these variable fluorescent
features is that they are scattered emission from the outburst of a
transient hard X-ray source, such as an X-ray binary or the
supermassive black hole \sgrastar. 

The luminosity of the source illuminating these reflection nebulae can 
be estimated from the optical depths of the scattering media 
coupled with the fluxes in the iron lines. 
The luminosity in the fluorescent Fe K$\alpha$ line is about 30\% of the 
flux absorbed by the photoelectric edge, and so the observed flux 
can be computed from
\begin{equation}
F_{K\alpha} = \frac{0.3}{4\pi D^2} \int_{E_0}^\infty  \frac{\mathcal{L}_{\rm X}(E)}{4\pi d^2} A 
\left[1 - e^{-\tau_{\rm Fe} [E/E_0]^{-3}} \right] dE,
\label{eq:fka}
\end{equation} 
where $D$ is the distance from the observer to the fluorescence
region, $\mathcal{L}_{\rm X}(E)$ is the specific luminosity of the
illuminating source as a function of energy (units of photons s$^{-1}$
keV$^{-1}$), $d$ is the distance between the source and the
fluorescing cloud, $A$ is the area of the cloud projected toward the
source, and $\tau_{\rm Fe}$ is the optical depth of the iron
photoelectric edge at an energy $E_0 = 7.1$ keV.

This equation can be simplified given the parameters of our nebulae.
The optical depth is given by $\tau_{\rm Fe} = N_{\rm H} \delta_{\rm
Fe} \sigma_{\rm Fe}$, where $\sigma_{\rm Fe} = 3.5 \times 10^{-20}$
cm$^{2}$ is the cross-section for photoelectric absorption at 7.1 keV,
and $\delta_{\rm Fe} = 3\times10^{-5}$ is the fractional abundance of
iron for solar abundances. If we take $N_{\rm H} \la 6\times10^{23}$
cm$^{-2}$ (Tab.~\ref{tab:spec}), the optical depth to iron
absorption is $\tau_{\rm Fe} \la 0.8$, and
$1 - e^{-\tau} \approx \tau$ with an error of $<$30\%.
Next, we take $D$=8 kpc, define $d_{10}$ to be normalized to 10 pc, re-write
$A$ as a solid angle $\Omega_1$ normalized to 1 arcmin$^2$
at 8 kpc, and write $\tau_{\rm Fe}$ in terms of the column depth $N_{\rm
H,23}$ in units of $10^{23}$ cm$^{-2}$ and the abundance of iron
relative to the solar value $\alpha_{\rm Fe}$.  Finally, we assume
that the spectrum of the illuminating source is a $\Gamma$=1.8 power 
law, $\mathcal{L}_{\rm X}(E) \propto E^{-1.8}$.  We then can integrate
eq.~\ref{eq:fka} over energy, and write the result in terms of the 
2--8 keV luminosity, normalized to $10^{38}$ \ergs, $L_{{\rm X},38}$. 
With these substitutions, we find:
\begin{equation}
F_{K\alpha} = 5\times10^{-5}~L_{{\rm X},38}~d_{10}^{-2}~N_{\rm H,23}~\alpha_{\rm Fe}~\Omega_1~{\rm ph~cm}^{-2}~{\rm s}^{-1}.
\end{equation}

If \sgrastar\ is the source of the photons that produced
the reflection features, and that the line-of-sight separation between
it and the cloud is equal to the projected separation (14 pc), 
$d \approx 20$ pc and an outburst with $L_{\rm X} \sim 10^{38}$ \ergs\ 
would have been required.\footnote{This
luminosity is larger than was reported in \citet{par04}, because a
mathematical error was made in computing the value in the earlier
paper.}  Such an outburst would have occurred 60 years ago, before the
advent of X-ray astronomy, and therefore would not have been
observed. The lack of similar reflection nebula at negative Galactic 
longitudes \citep{par04} implies that those clouds do not lie within
the light front of the outburst. 

 Alternatively, if the bright transient that was in outburst
in 2004 produced the fluorescent features, $d \approx 7$ pc
and $L_{\rm X} \sim 10^{37}$ \ergs. Although this is 100 times
brighter than the outbursts in 2002 and 2004 \citep{por05, m-trans},
this outburst would have occurred 20 years ago, before the
advent of both \chandra\ and the wide-field X-ray monitors on {\it
BeppoSAX} and the {\it Rossi X-ray Timing Explorer}, so it too could
have been missed.  However, it is uncommon for the outburst of an X-ray
binary to persist at this luminosity for 2--3 years 
\citep{csl97}\footnote{See also 
{\tt http://xte.mit.edu/ASM\_lc.html}.}, so we prefer the 
hypothesis that the source of the outburst was \sgrastar. 


The time evolution of a reflection nebula has been explored in 
detail by \citet{sc98}, assuming target material with a spherical
geometry. However, the observed changes in the morphology in Figure
\ref{fig:images} are significantly more complex than they considered. 
For instance, in Feature 1 the vertical filament
$\approx$2 parsec long fades over the course of 3 years. The fact that
the light-travel time along the feature is longer than the interval
over which it fades implies that the light front propagated 
perpendicular to the filament, and had width that is similar to that
of the filament (0.3 pc). The fact that the whole filament 
dimmed, with very little brightening adjacent
to most of its length (aside from knots in the north on either side of
the filament) implies that the X-rays were no longer
propagating through dense material in 2004 and 2005. Therefore, the 
morphology of the iron features traces the structure of the 
clouds, revealing that they contain filamentary 
features a few tenths of a parsec wide and a couple parsecs long.

\section{Conclusions}

The discovery of changes in the intensity and morphology of
fluorescent iron nebulae near the Galactic center demonstrates
that they are reflection nebulae. The most likely source of the
reflected X-rays is an outburst of \sgrastar\ with a luminosity 
of at least $10^{38}$ \ergs, which would have occurred about 
50 years ago and lasted 2--3 years. If a similar outburst had
occurred since then, it would have had a good chance of being
detected by an X-ray instrument, or would have produced similar 
fluorescent features in other, nearby clouds. Since no other, 
long outburst of \sgrastar\ has been reported, and no fluorescent 
features have been identified closer to \sgrastar\ \citep{par04},
the duty cycle of bright, year-long outbursts from \sgrastar\ 
must be $\la$5\%. 


These fluorescent features also reveal filamentary 
structures within the dense regions of the molecular clouds.
Interferometric millimeter observations would resolve clouds
on a similar spatial scale, and additionally provide velocity 
information. Combining the cloud velocities and the 
three-dimensional spatial information afforded by the evolution of
the reflection nebulae would allow one to construct a physical model
of these turbulent molecular clouds. This is a necessary step 
toward modeling star formation in the hot, turbulent
environment of the Galactic center. 

\acknowledgments
This work was funded by NASA through the {\it Chandra}
X-ray Observatory.

\end{document}